# *Suggest an Aspect-Oriented Design Approach for UML Communication Diagram*

Mohammed F.Nather
Software Engineering Dept.
University of Mosul
Mosul, Iraq
mhmd.software@yahoo.com

Dr.Nada N.Saleem
Software Engineering Dept.
University of Mosul
Mosul, Iraq
Na_ni_s@yahoo.com

***Abstract-* More and more works are done on the design of the Unified Modeling Language (UML) which is designed to help us for modeling effective object oriented software , Existing Object-Oriented design methods are not mature enough to capture non-functional requirement such as concurrency, fault tolerance , distribution and persistence of a software approach. Our approach proposed to use aspect-oriented software development (AOSD) mechanisms to solve the issues for interactions of the communication diagram in UML that support only the Object-Oriented mechanisms ,thus AOSD allow to design programs that are out of reach of strict Object-Orientation and could possibly improve the structures and implementations.**

***Keywords-Aspect-Oriented Programming (AOP), Unified Modeling Language (UML) Aspect Oriented Software Development (AOSD), Software Engineering (SE), Separation of Concerns (SoC)***

## I. Introduction

Aspect-Oriented Software Development (ASOD) has arisen as an approach that supports a better Separation of Concerns (SoC) and more adequately reflects the way developers think about the system [3]. Essentially, AOSD introduces a unit of modular implementation – the aspect – which has been typically used to encapsulate crosscutting concerns in software systems (i.e., concerns that are spread across or tangled with other concerns). Modularity, maintainability, and facility to write software can be achieved with AOP [9]. At the early stages of software development, however, concerns strongly overlap with broadly scoped qualities such as performance, reliability, and security with relationships that are often of a qualitative nature. These relationships go well beyond common composition rules such as before, after, around, concurrent, and interleaved . For example, a security concern surely impacts negatively a performance concern, because more resources are required for security features such as encryption, authentication, or access control, but just how much on the other hand, a performance concern can affect negatively a security concern if the performance concern caches results, which must then be protected. These are examples of qualitative interactions. Interactions are one of the most interesting kinds of relationships between concerns, because they describe potentially undesirable impact of one concern on another. Interactions have been studied extensively in the telecommunications domain but are applicable to many other domains and applications. Interactions manifest themselves in aspect-oriented models, when multiple aspects affect the same elements in the base [11].

## II. ASPECT-ORIENTED PROGRAMMING

AOP is merging discipline in Software Engineering, aimed at modularizing crosscutting concerns by encapsulating replicated, scattered, and tangled code in Aspects , Aspect-Oriented Software development (ASOD) is a new technique to support separation of concerns in software development [1], Examples of crosscutting concerns include tracing, logging, caching, resource pooling and so on. The techniques of AOSD make it possible to modularize crosscutting aspects of a system. Like objects in object-oriented software development, aspects in AOSD may arise at a stage of the software life cycle, including , requirements, specification, design, implementation, etc [4] .

### *A. Aspectj*

AOP has been implemented in different languages, among them Smalltalk and Java . The Java implementation of AOP is called AspectJ (TM) and has been created at Xerox PARC [14]. AspectJ is an aspect-oriented extension to Java. The language is fully compatible with pure Java. However, it introduces new kinds of structures and new keywords to write aspects [6], which adds AOP capabilities to Java.

AspectJ includes following attributes:

- Aspect is a module for handling crosscutting concerns .Aspects are defined in terms of pointcuts, advice, and introduction .Aspects are reusable and inheritable .
- Joinpoint is a Point of execution of java program. it further includes constructor call , method call , exception handler execution .
- pointcut is a predicate that matches join points.
- advice is code that is executed at a pointcut ,which is an action taken by an aspect at a particular joinpoint,

different types of advice include ″around″, ″before″ and “after” advice.

- Weaving linking aspects with other application type or objects to create an advised object,this can be done at compile time (using the AspectJ compiler, for exam ple), load time, or at runtime.
- Target object is an object being advised by one or more aspects also referred to as the advised object [13].

### *B . ASPECT-ORIENTED MODELING AND DESIGN*

AOM aims at supporting separation of crosscutting concerns at the modeling level, with the purpose of enhancing productivity, quality and reusability through the encapsulation of requirements that cut across software components. Aspect-oriented modeling (AOM) is therefore of great interest , which involves identifying, analyzing, managing, and representing crosscutting concerns. Zooming into design-level approaches Aspect-Oriented Modeling (AOM) provides support for separating concerns at the design level and has the potential to effectively tackle the complexity of developing software that deals with interdependent concerns[16]. Aspect-oriented design has the same objectives as any software design activity, i.e. characterizing and specifying the behavior and structure of the software system. Its unique contribution to software design lies in the fact that concerns that are necessarily scattered and tangled in more traditional approaches can be modularized, even with proper education understanding crosscutting concerns can be difficult without proper support for visualizing both static structure and the dynamic flow of a program. So the languages which implement AOP must have the facility to support the visualizing of crosscutting concerns[15] , this section describes our motivation to support aspects in the design level and describe them in XMI. Then, we outline the underlying technologies, UML, XML.

### *C. Benefits of Capturing Aspects in the Design Phase*

Aspects can be identified at the requirement, design and implementation phases, though the inter-component tangling tends to occur at the implementation/coding phase. When aspects are identified, or emergent, at the implementation phase, developers often add or change aspects manually and maintain them in the source code level. Few methods have been proposed for expressing aspects in the design level. Supporting aspects at the design phase streamlines the process of aspect-oriented development by facilitating:

- Documentation and Learning: Supporting an aspect as a design construct allows developers to recognize it in the upper level of abstraction at the earlier stage of development process. Aspect designers and people learning aspect-oriented software can learn and document aspect models in more intuitive way. For example, they can visualize aspect models using a CASE tool that supports visual modeling [15][8].
- Reuse of aspect: The ease of documentation and learning leverages the reuse of aspect information, how an aspect is designed and how it is intended to affect to classes. It’s easy to imagine more sophisticated ways of using aspects, such as aspect-aware CASE tools, hyperlinked documents and pattern catalogues that collects well-known and feasible aspects. These would increase the reusability of aspect-based design [15],[8].

## III. UNIFIED MODELING LANGUAGE (UML)

The Unified Modeling Language (UML) is a general-purpose graphical object-oriented modeling language that is designed to visualize, specify, construct and document software systems in both structural and behavioral aspects. UML is intended to be a common way of capturing and expressing relationships and behaviors in a notation that's easy to learn and efficient to write [10]. UML Is a design language most accepted in software engineering, and is considered as a standard. It includes many useful ideas and concepts that have their roots in various individual methods and theories. UML provides numerous modeling techniques, including several types of diagrams, model elements, notation and guidelines. These techniques can be used in various ways to model different characteristics of a software system [4], UML 2 describes 13 official diagram types which classified into structural and behavioral diagrams, see figure 1 [2].

### *A. interaction diagrams*

Classified into four types of diagrams (sequence diagram, communication diagrams, interaction overview diagrams, timing diagrams) helps you accurately model how the parts that make up your system interact [7].

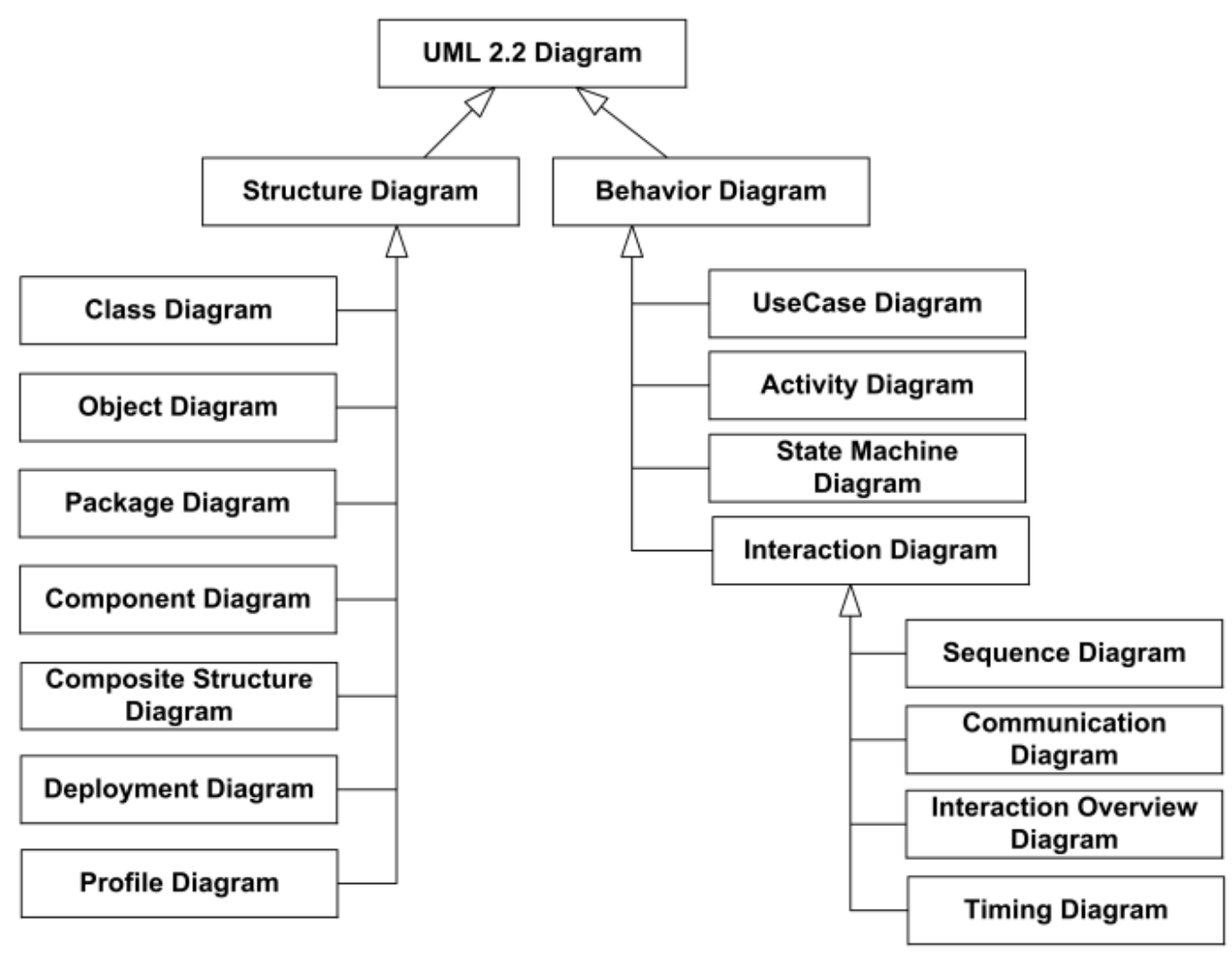

**Figure 1: 1UML 2.0 Diagrams**

### *B. Communication Diagram*

A kind of interaction diagram, emphasize the data links between the various participants in the interaction. Instead of drawing each participant as a lifeline and showing the sequence of messages by vertical direction as the sequence diagrams does, the communication diagram allows free placement of participants, allows you to draw links to show how the participants connect, and use numbering to show the sequence of messages. In UML 1 .x, these diagrams were called collaboration diagrams. This name stuck well, and I suspect that it will [r10].

## IV. XML METADATA INTERCHANGE (XMI)

XMI is a standard that enables you to express your objects using Extensible Markup Language (XML), XMI provides a standard representation of objects in XML, enabling the effective exchange of objects using XML. It is intended to provide a standard way for programmers and other users to exchange information about metadata (essentially, information about what a set of data consists of and how it is organized). Specifically, XMI is intended to help programmers using the Unified Modeling Language (UML) with different languages and development tools to exchange their data models with each other [5].

## V. OUR WORK

in this section our approach explained by the constructed software engineering tool, which produced to extend UML to support designing the communication diagram with the concepts of Aspect-Oriented modeling (AOM) .The tool input is reading XMI file obtained from other tools that describe the problem with communication diagram in Object-Oriented like EA, then parsing the XMI file to get the useful information . after that the Aspect-Oriented approach analyze the information and identify the crosscutting of the non-functional concerns , finally designing the communication diagrams with the notation of aspect-oriented and generate code for the aspect/class and the other classes for the functional concerns see figure (2) for the processing .

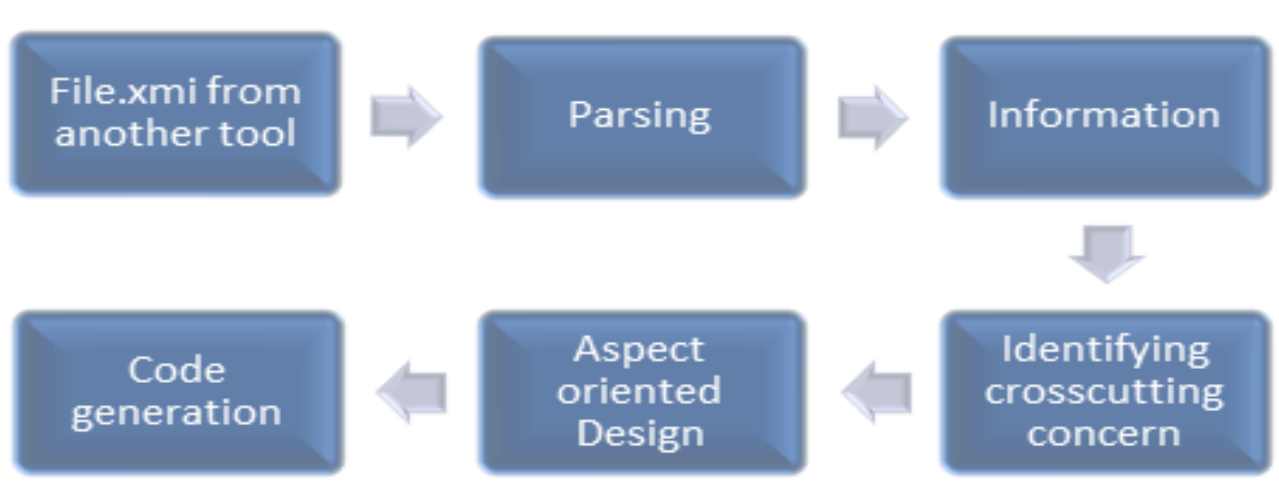

**Figure 2: work processing**

### *A. Parsing process*

When dealing with XMI programmatically, one of the first things you have to do is take an XML document and parse it, I use Simple application program interface (API) for XML (SAX) parsing. As the document is parsed, the data becomes available to the application, see figure (3) for the XMI 1.1 generated form enterprise architect (EA) tool from sparx system products, this figure show the description of a message between two objects.

### *B. Table of information*

After the parsing process the data is available for processing and saved in tables, each row contain the properties of each message between two objects, ( the name of the message , object sender , object receiver , class sender , class receiver , the classes are functional or non-functional , the sequence of the message , and the diagram to which the message back) See figure (4).

### *C. Identifing crosscutting and AOD*

In object-oriented programming (OOP), ones usually captures the main concerns of a problem (e.g. for a banking application, money withdrawal at an ATM, funds transfer,...) and then encapsulates them into classes. The problem is that some “secondary”, non-business operations are needed (non-functional, i.e. logging, authentication, transactions), these operations are shared by several classes of a system, and each of them has to code their firing. The main idea behind AOP is to consider these operations as crosscutting aspects and to group them into separate modules, which are automatically called by the system. This new approach helps building clean software architectures and completes the toolbox of software engineers, by capturing the crosscutting depending on the number of repetition of the method call (which is the message in communication diagram) from the functional concern to non-functional concern, then compare the number with the specified threshold. The classes for the non-functional concern suggested to become an aspect/classes, then redesign each diagram that specify each operation in the system .(the problem as an input in communication diagram within OOD) , see figure(5) , and figure(6) that show how Aspect are captured and represented with red color within the notation of Aspect-Oriented design . Finally applying the fan-in and fan-out metrics to the problem in OOD to find the coupling of each class, and then apply the same metrics to the AOD.

### *D. Code generation*

From figure (6) the tool can generate code for each aspect/class or other classes by double clicking on each one, see figure (7).

## VI. CONCLUSION AND FUTURE WORK

We have seen in this paper how AOP addresses non-functional requirements (crosscutting concern) using java through the constructed tool to support the work. The aim for having these crosscutting concerns being implemented separately from the core concern to bring clearly the specific implementation of our aspects to maximum understandability, which leads to better maintenance, and reusability. By applying fan-in and fan-out metrics on the problem in OOP and after redesigning the problem with AOP we found the coupling in decreased between classes and the structure is improved. For future work we suggest to use artificial intelligence algorithms to identify and capture the crosscutting concern.

```
<UML:Message name="block user" xmi.id="EAID_B479EEB2_DC42_43fe_BC24_6FAA8A70B5E8" visibility="public"
sender="EAID_0DF2BC6D_B53B_4536_BC6D_2E5B175A905C" receiver="EAID_4E2A843C_3F57_4b79_8465_C696EF958116"
collaboration="EAID_1CE1F0B8_5AAD_4303_A642_3AA6AFCC51EC">
    <UML:ModelElement.taggedValue>
        <UML:TaggedValue tag="message_link" value="EAID_1CE1F0B8_5AAD_4303_A642_3AA6AFCC51EC"/>
        <UML:TaggedValue tag="style" value="1"/>
        <UML:TaggedValue tag="ea_type" value="Collaboration"/>
        <UML:TaggedValue tag="direction" value="Source -&gt; Destination"/>
        <UML:TaggedValue tag="linemode" value="1"/>
        <UML:TaggedValue tag="linecolor" value="-1"/>
        <UML:TaggedValue tag="linewidth" value="0"/>
        <UML:TaggedValue tag="seqno" value="15"/>
        <UML:TaggedValue tag="headStyle" value="0"/>
        <UML:TaggedValue tag="lineStyle" value="1"/>
        <UML:TaggedValue tag="conditional" value="any missuse"/>
        <UML:TaggedValue tag="privatedata1" value="Synchronous"/>
        <UML:TaggedValue tag="privatedata3" value="Call"/>
        <UML:TaggedValue tag="privatedata4" value="5.2.1"/>
        <UML:TaggedValue tag="ea_localid" value="190"/>
        <UML:TaggedValue tag="ea_sourceName" value="monitoring "/>
        <UML:TaggedValue tag="ea_targetName" value="h-page"/>
        <UML:TaggedValue tag="ea_sourceType" value="Object"/>
        <UML:TaggedValue tag="ea_targetType" value="Object"/>
        <UML:TaggedValue tag="ea_sourceID" value="63"/>
        <UML:TaggedValue tag="ea_targetID" value="51"/>
        <UML:TaggedValue tag="src_visibility" value="Public"/>
        <UML:TaggedValue tag="src_isOrdered" value="false"/>
        <UML:TaggedValue tag="src_targetScope" value="instance"/>
        <UML:TaggedValue tag="src_changeable" value="none"/>
        <UML:TaggedValue tag="src_isNavigable" value="false"/>
        <UML:TaggedValue tag="src_containment" value="Unspecified"/>
        <UML:TaggedValue tag="dst_visibility" value="Public"/>
        <UML:TaggedValue tag="dst_aggregation" value="0"/>
        <UML:TaggedValue tag="dst_isOrdered" value="false"/>
        <UML:TaggedValue tag="dst_targetScope" value="instance"/>
        <UML:TaggedValue tag="dst_changeable" value="none"/>
        <UML:TaggedValue tag="dst_isNavigable" value="true"/>
        <UML:TaggedValue tag="dst_containment" value="Unspecified"/>
        <UML:TaggedValue tag="diagram" value="EAID_62A38B1B_58D5_4985_A568_43A9024B9734"/>
        <UML:TaggedValue tag="lt" value="5.2.1: [any missuse]:block user()"/>
    </UML:ModelElement.taggedValue>
```

**Figure 3: sample of XMI**

MyTool

Open XML pack...

Parsing

print

Choose Diagram :

draw

| Message Name | Object Sender | Class Sender | Concern Type ... | Object Receiver | Class Receiver | Concern Type ... | Message Sequ... | Diagram Name | Repetition |
|---|---|---|---|---|---|---|---|---|---|
| access | user | customer | functional req... | h-page | home page | functional req... | 1: access() | cheque service | |
| click login | h-page | home page | functional req... | l-page | login page | non functional... | 1.1: click login() | cheque service | 6 |
| display login | l-page | login page | non functional... | user | customer | functional req... | 1.2: display lo... | cheque service | |
| enter | user | customer | functional req... | l-page | login page | non functional... | 2: enter(user... | cheque service | 6 |
| verify | l-page | login page | non functional... | db | account data ... | functional req... | 2.1: verify(use... | cheque service | |
| valid user | db | account data ... | functional req... | om-page | option menu p... | functional req... | 2.2: valid user() | cheque service | |
| display option... | om-page | option menu p... | functional req... | user | customer | functional req... | 2.3: display o... | cheque service | |
| check behavior | h-page | home page | functional req... | monitoring | security | non functional... | 3: behave= c... | cheque service | 5 |
| recording use... | monitoring | security | non functional... | db | account data ... | functional req... | 3.1: recording... | cheque service | |
| select cheque... | user | customer | functional req... | om-page | option menu p... | functional req... | 4: select cheq... | cheque service | |
| display chequ... | om-page | option menu p... | functional req... | db | account data ... | functional req... | 4.1: display= ... | cheque service | |
| check user st... | db | account data ... | functional req... | monitoring | security | non functional... | 4.2: check us... | cheque service | 4 |
| block user | monitoring | security | non functional... | h-page | home page | functional req... | 4.2.1: [any mi... | cheque service | |
| logout | h-page | home page | functional req... | user | customer | functional req... | 4.2.2: logout() | cheque service | |
| view cheque ... | om-page | option menu p... | functional req... | user | customer | functional req... | 4.3: view che... | cheque service | |
| access | user | customer | functional req... | h-page | home page | functional req... | 1: access() | logging | |
| click login | h-page | home page | functional req... | l-page | login page | non functional... | 1.1: click login() | logging | 6 |
| display login | l-page | login page | non functional... | user | customer | functional req... | 1.2: display lo... | logging | |
| enter | user | customer | functional req... | l-page | login page | non functional... | 2: enter(user... | logging | 6 |
| verify | l-page | login page | non functional... | db | account data ... | functional req... | 2.1: verify(use... | logging | |
| valid user | db | account data ... | functional req... | om-page | option menu p... | functional req... | 2.2: valid user() | logging | |
| display option... | om-page | option menu p... | functional req... | user | customer | functional req... | 2.3: display o... | logging | |
| access | user | customer | functional req... | h-page | home page | functional req... | 1: access() | pay bills | |

Package Path : C:\Users\hp_dv3\Desktop\final.xml

**Figure 4: Table showing the information after parsing**

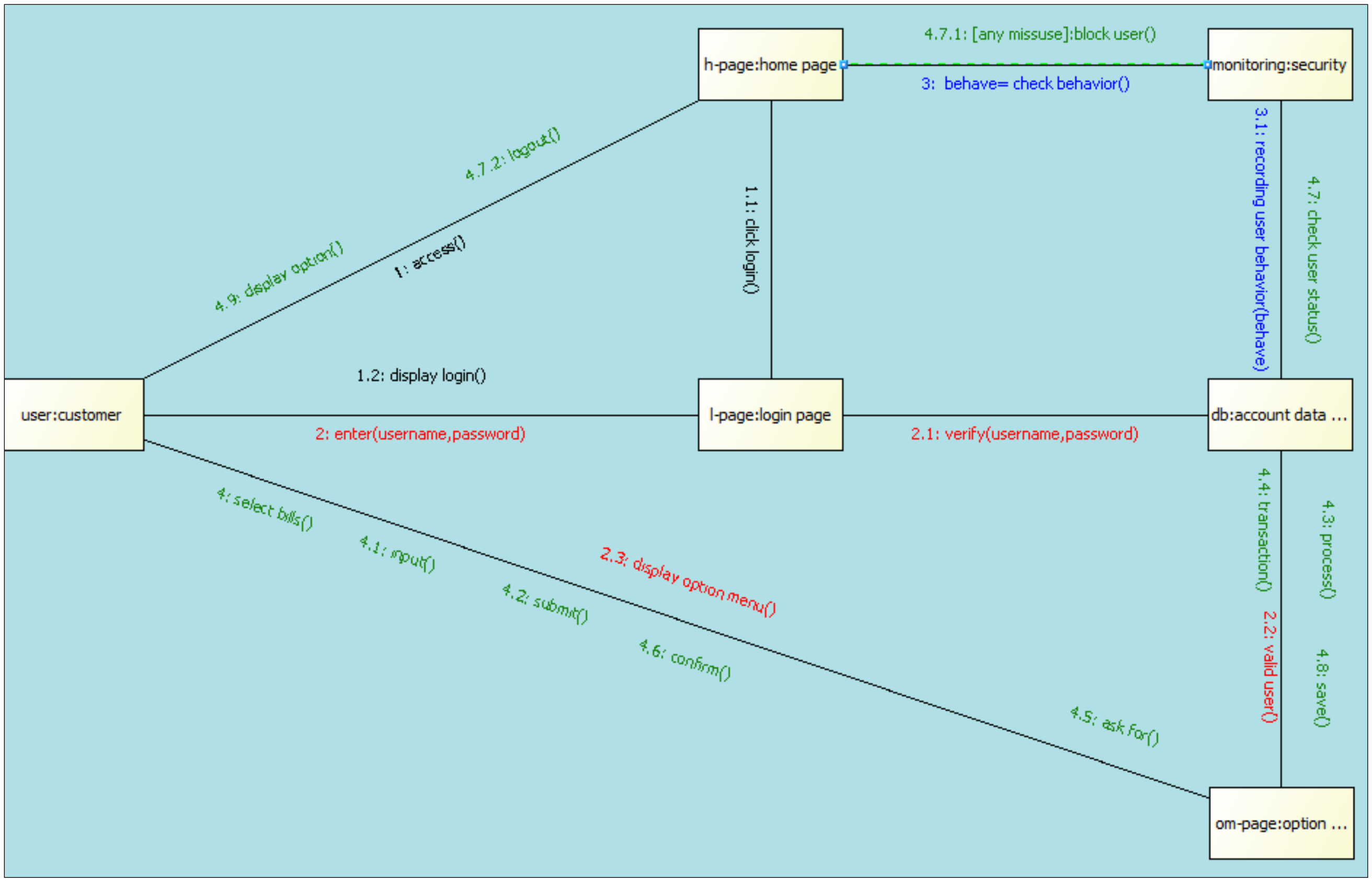

**Figure 5: communication diagram in OOP**

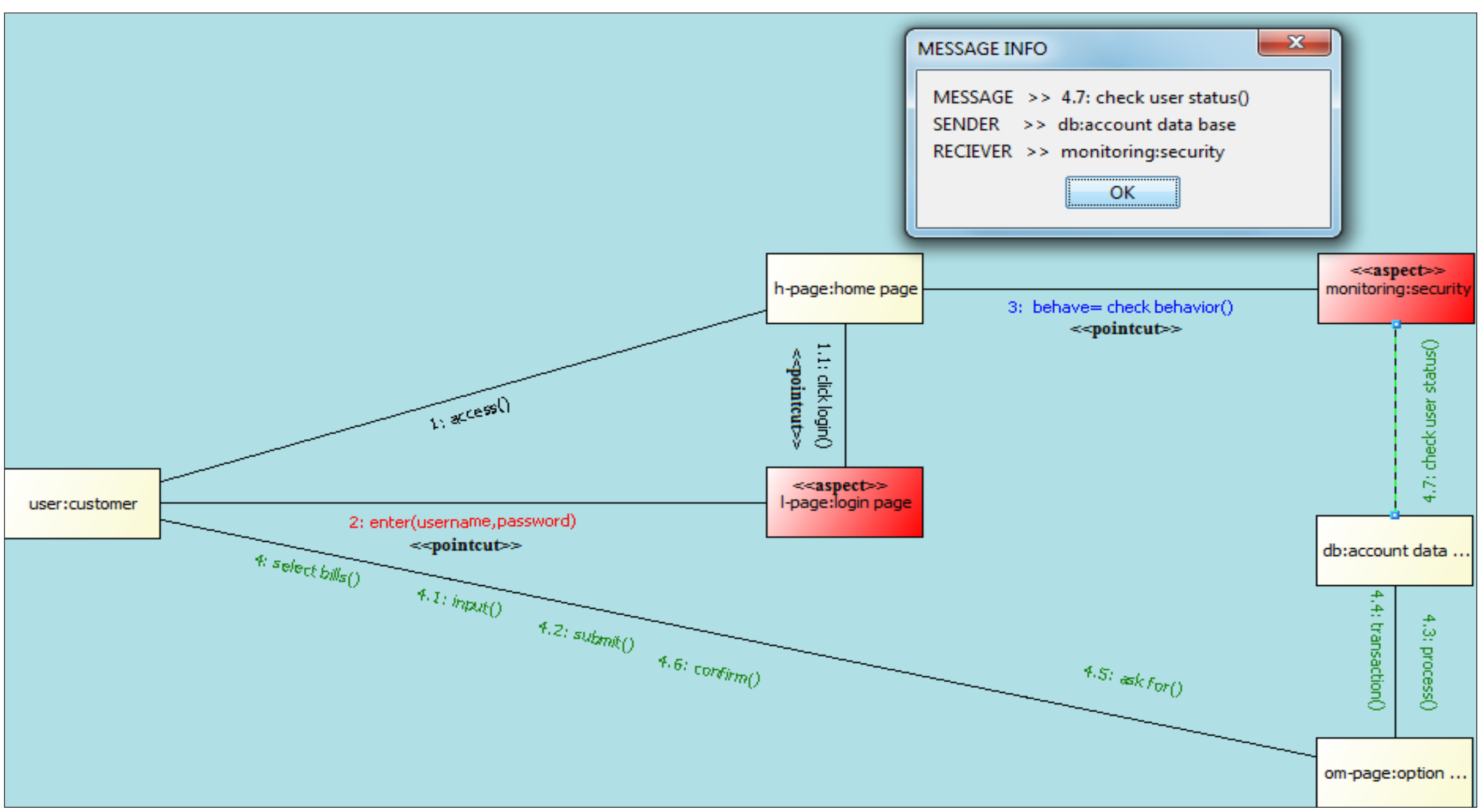

**Figure 6: Applying AOD approach**

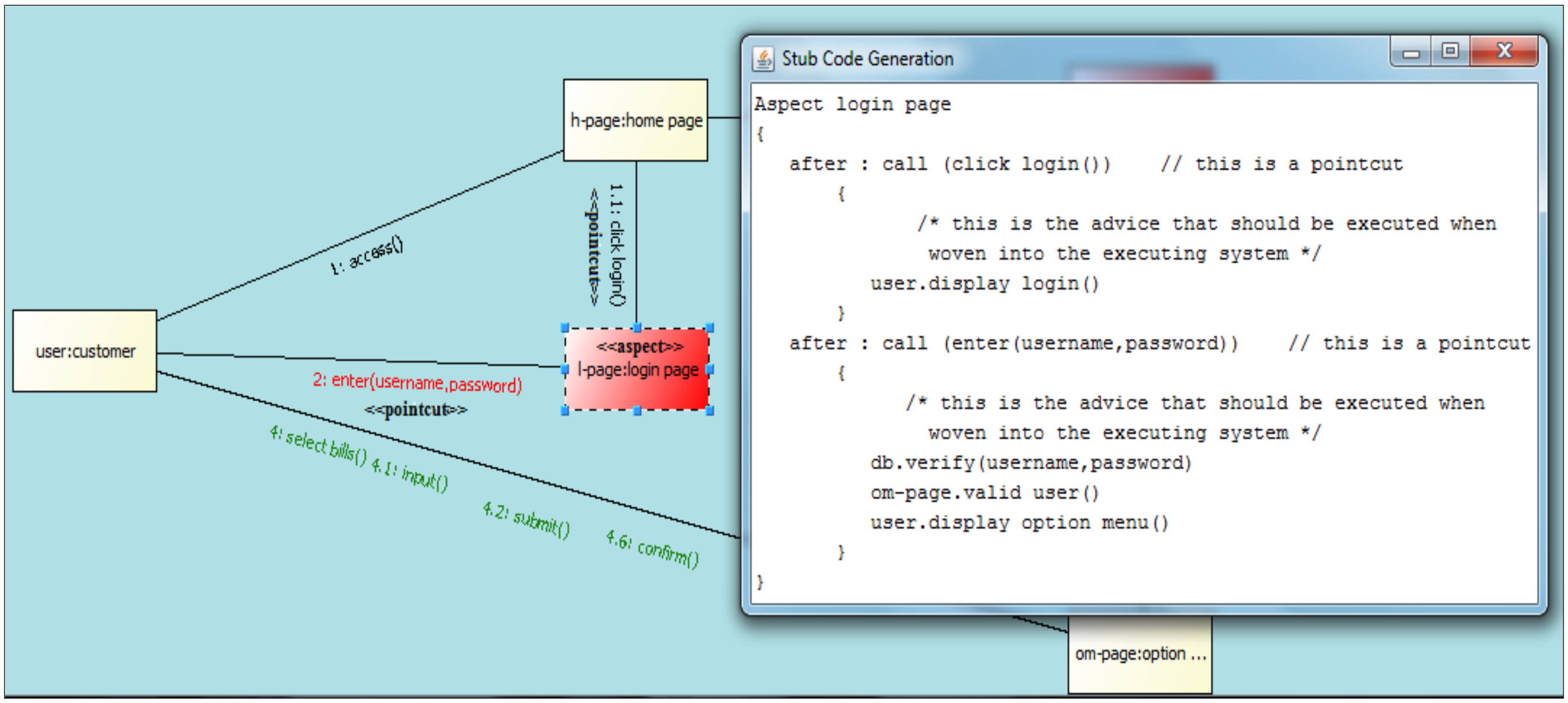

**Figure7: Code generation from AOD**